\documentclass[conference]{IEEEtran}

\IEEEoverridecommandlockouts
\usepackage{graphicx}
\usepackage{amsmath,amssymb,amsthm}
\usepackage{textcomp}
\usepackage{cite}
\usepackage{mathtools}
\usepackage{subfig}
\usepackage[nolist]{acronym}
\usepackage[hidelinks]{hyperref}
\usepackage{cleveref}
\usepackage{relsize}
\usepackage{float}
\usepackage[ruled,vlined]{algorithm2e}
\usepackage{tabularx}
\usepackage{booktabs}
\usepackage{mathtools}
\usepackage{enumitem}
\usepackage{pgf,tikz}
\usepackage{pgfplots}
\usetikzlibrary{arrows,positioning} 
\usetikzlibrary{decorations.text}
\usetikzlibrary{arrows.meta}
\usepgfplotslibrary{external}
\tikzsetexternalprefix{external_figs/}
\usepackage{contour}

\graphicspath{{./graphics/}}

\theoremstyle{definition}

\theoremstyle{definition}

\theoremstyle{definition}

\newcommand{\argmax}{\operatornamewithlimits{arg\,max}}

\definecolor{blue33D}{RGB}{0, 0, 175}

\newcommand{\legendfont}{\fontsize{9pt}{9pt}\selectfont}

\newcommand{\labelfont}{\fontsize{10pt}{10pt}\selectfont}

\definecolor{errorBarC}{RGB}{150,150,150}
\definecolor{c1}{RGB}{215, 196, 255}
\definecolor{c2}{RGB}{89,51,84}
\definecolor{cgrid}{RGB}{215,215,215}
\definecolor{cgrid2}{RGB}{215,215,215}
\definecolor{dark01}{RGB}{75, 75, 75}
\definecolor{greenC1}{RGB}{13, 128, 82}
\definecolor{redC1}{RGB}{209, 6, 60}
\definecolor{blueC1}{RGB}{6, 67, 209}
\definecolor{c1}{RGB}{209, 6, 60}
\definecolor{c2}{RGB}{6, 67, 209}
\definecolor{c3}{RGB}{13, 128, 82}
\definecolor{c4}{RGB}{126, 47, 142}
\definecolor{c5}{RGB}{222, 125, 0}
\definecolor{c6}{RGB}{20, 43, 140}
\definecolor{c7}{RGB}{162, 20, 47}

\tikzset{
	gridc/.style= {dotted, cgrid2},
	linew/.style= {line width=1pt, mark options={solid}},
	linew2/.style= {line width=1pt, dashed, dash pattern=on 5pt off 4pt, mark options={solid}},
	linew3/.style= {line width=1pt, dashed, dash pattern=on 1pt off 2pt, mark options={solid}},
	marksz/.style= {mark options={scale=1, fill opacity=1, solid, line width=0.5}},
	marksz2/.style= {mark options={scale=1.4, fill opacity=1, solid, rotate=180, line width=0.5}},
	marksz3/.style= {mark options={scale=1, fill=white, fill opacity=0.5, solid, line width=0.5}},
	errorlinemarkSty/.style = {mark size=0.25pt, solid, color=errorBarC},
	errormarkSty/.style = {rotate=90, mark size=0pt, solid, color=errorBarC},
	errorbarSty/.style  = {solid, line width=1pt, opacity=0.75, color=errorBarC, cap=round},
	invisible/.style={opacity=0},
	visible on/.style={alt={#1{}{invisible}}},
	alt/.code args={<#1>#2#3}{%
		\alt<#1>{\pgfkeysalso{#2}}{\pgfkeysalso{#3}} 
	},
}

\pgfplotsset{
	axisSetup/.style= {axis x line=bottom, axis y line=left, tick align=inside, axis line style={-, line width=1.25pt, color=dark01}},
	short Legend0/.style={%
		legend image code/.code={
			\draw[##1,line width=1pt] plot coordinates {(0pt,0pt) (15pt,0pt)};
		}
	},
	short Legend1/.style={%
		legend image code/.code={
			\draw[##1,line width=1pt] plot coordinates {(0pt,0pt) (15pt,0pt)};
		}
	},
	short Legend2/.style={%
		legend image code/.code={
			\draw[##1,linew] plot coordinates {(0pt,0pt) (9pt,0pt)};
			\draw[##1,linew, dashed] plot coordinates {(0pt,2pt) (9pt,2pt)};
			\draw[##1,mark=o, marksz, linew] plot coordinates {(-4pt,1pt)};
		}
	},
	short Legend3/.style={%
		legend image code/.code={
			\draw[##1,linew] plot coordinates {(0pt,0pt) (9pt,0pt)};
			\draw[##1,linew, dashed] plot coordinates {(0pt,2pt) (9pt,2pt)};
			\draw[##1,mark=triangle, linew, marksz2] plot coordinates {(-4pt,1pt)};
		}
	},
	ylabel right/.style={
		after end axis/.append code={
			\node [rotate=90, anchor=north] at (rel axis cs:1,0.5) {#1};
		}   
	}
}

\newif\ifshowtikz
\showtikztrue

\let\oldtikzpicture\tikzpicture
\let\oldendtikzpicture\endtikzpicture

\renewenvironment{tikzpicture}{%
	\ifshowtikz\expandafter\oldtikzpicture%
	\else\comment%
	\fi
}{%
	\ifshowtikz\oldendtikzpicture%
	\else\endcomment%
	\fi
}

\begin{document}

\author{Bashar Tahir, Stefan Schwarz, and Markus Rupp \\
	
	\thanks{Bashar Tahir and Stefan Schwarz are with the Christian Doppler Laboratory for Dependable Wireless Connectivity for the Society in Motion. The financial support by the Austrian Federal Ministry for Digital and Economic Affairs and the National Foundation for Research, Technology and Development is gratefully acknowledged.}

	Institute of Telecommunications, Technische Universit\"{a}t Wien, Vienna, Austria 
}

\title{RIS-Assisted Code-Domain MIMO-NOMA}

\maketitle
\begin{abstract}
We consider the combination of uplink code-domain non-orthogonal multiple access (NOMA) with massive multiple-input multiple-output (MIMO) and reconfigurable intelligent surfaces (RISs). We assume a setup in which the base station (BS) is capable of forming beams towards the RISs under line-of-sight conditions, and where each RIS is covering a cluster of users. In order to support multi-user transmissions within a cluster, code-domain NOMA via spreading is utilized. We investigate the optimization of the RIS phase-shifts such that a large number of users is supported. As it turns out, it is a coupled optimization problem that depends on the detection order under interference cancellation and the applied filtering at the BS. We propose to decouple those variables by using sum-rate optimized phase-shifts as the initial solution, allowing us to obtain a decoupled estimate of those variables. Then, in order to determine the final phase-shifts, the problem is relaxed into a semidefinite program that can be solved efficiently via convex optimization algorithms. Simulation results show the effectiveness of our approach in improving the detectability of the users.
\end{abstract}


\IEEEpeerreviewmaketitle
\vspace{-1mm}
\section{Introduction}
\vspace{-0.5mm}
\Acfp{RIS} have emerged as a promising technology for \ac{B5G} wireless communication, enabling operation with high spectral and energy efficiency \cite{Renzo19, Wu19}. Consisting of configurable nearly-passive elements, those surfaces are capable of altering the propagation of the electromagnetic waves impinged on them, allowing them to perform passive beamforming of the waves from and to a certain point, suppress interference, extend the coverage area, etc \cite{Wu20, Basar19}. Combining \acp{RIS} with \ac{NOMA} has been the focus of many works, such as \cite{Fu19, Ding20a, Yang20, Mu20, Sena20}, showing potential gains in terms of the energy efficiency, sum-rate, and outage performance. In \ac{NOMA}, multiple \acp{UE} occupy the same time-frequency resources,  which may lead to a higher spectral efficiency, lower access latency, improved user fairness \cite{Dai18, Ding17}. So far, those works focused on pure power-domain \ac{NOMA}, and on the optimization of the \ac{RIS} phase-shifts under \ac{NOMA} \ac{IC}. Little attention has been given to code-domain \ac{NOMA}, where on top of the power-domain superposition, the \acp{UE} transmit with code-domain signatures (e.g., short spreading sequences), which permits interference suppression at the receiver via code-domain processing \cite{Cai18, Wu18}. Such an interference suppression capability allows for a large number of simultaneous transmissions; combined with massive \ac{MIMO}, this can help enabling massive connectivity.

\begin{figure}
	\centering
	\begin{center}
		\begin{tikzpicture}
			\node[,] (image) at (0,0) {\includegraphics[width=0.75\linewidth]{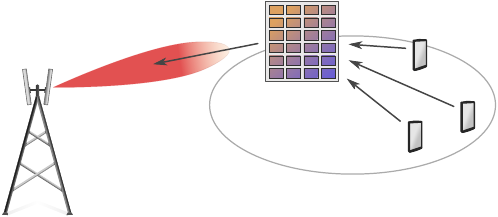}};
			\node[,] at (0.95, -0.45) {\legendfont Target cluster};
		\end{tikzpicture}
	\vspace{-3mm}
	\end{center}
	\caption{The cluster-based RIS-assisted NOMA uplink.}
	\label{fig:00}
	\vspace{-5mm}
\end{figure}

We investigate here the combination of uplink code-domain \ac{NOMA} with \acp{RIS}, in the context of a cluster-based massive \ac{MIMO} deployment. Since it is likely that those surfaces would be deployed at rooftops, we make the assumption that each cluster is served by a \ac{RIS} having a strong \ac{LOS} connection to the \ac{BS}. The \ac{BS} forms beams towards the clusters' \acp{RIS}, allowing to simultaneously boost the received power of the target cluster and suppress intercluster interference, as depicted in \Cref{fig:00}. In order to support massive connectivity, code-domain \ac{NOMA} via short spreading is employed in each cluster. At the \ac{BS}, and after spatial filtering, \ac{MMSE}-\ac{IC} detection is carried out to detect the \ac{NOMA} \acp{UE}. The question then is, how to configure the \ac{RIS} such that a large number of \acp{UE} is supported? As we will see later, it is a coupled optimization problem that depends on multiple variables, such as the detection order under \ac{IC}, and the applied \ac{MMSE} filters. To this end, we propose to obtain a decoupled estimate of those variables by utilizing sum-rate optimized phase-shifts as an initial solution. We then find the final shifts by a \ac{SDP} relaxation of the optimization problem, which can be solved efficiently via the framework of convex optimization. The simulation results show that our proposed approach can substantially improve the detection performance.
\vspace{-5mm}
\section{System Model}
We consider a cluster of $K$ single-antenna \acp{UE} communicating with an $N_r$-antennas \ac{BS} through an $N_s$-elements \ac{RIS}. Due to blockage, we assume the communication to take place primarily through the \ac{RIS}, and therefore we drop the direct paths between the \acp{UE} and the \ac{BS}. We will further justify this assumption later. Each \ac{UE} transmits using a short spreading signature of length $L$, assigned to them by the \ac{BS} based on a non-orthogonal spreading codebook. Assuming the channel is flat over the spreading interval (valid for small $L$), the received signal at the \ac{BS}, $\mathbf{y} \in \mathbb{C}^{N_rL \times 1}$, is given by
\begin{align}\label{eq:01}
	\mathbf{y} = \sum_{k = 1}^{K} \sqrt{\ell_{\textrm{BS}} \ell_{h_k} P_k L}\, \big(\mathbf{H}_{\textrm{BS}}\mathbf{\Phi}\mathbf{h}_{k} \otimes \mathbf{s}_k\big)x_k + \mathbf{z} + \mathbf{n},
\end{align}
where $\ell_{\textrm{BS}}$ and $\ell_{h_k}$ are the pathlosses of the \ac{BS}-\ac{RIS} and $k^{\text{th}}$ \ac{RIS}-\ac{UE} channels, respectively, and $P_k$ is the transmit power of the $k^{\text{th}}$ \ac{UE}. The quantities $\mathbf{H}_{\textrm{BS}} \in \mathbb{C}^{N_r\times N_s}$ and $\mathbf{h}_{k} \in \mathbb{C}^{N_s\times 1}$ represent the small-scale fading of the \ac{BS}-\ac{RIS} and $k^{\text{th}}$ \ac{RIS}-\ac{UE} channels, respectively. The matrix $\mathbf{\Phi} \in \mathbb{C}^{N_s\times N_s} = \textrm{diag}(e^{j\phi_1},\,e^{j\phi_2},\,\dots,\, e^{j\phi_{N_s}})$ is the phase-shift matrix applied at the \ac{RIS}, where $\phi_n$ is the phase-shift applied at the $n^{\text{th}}$ element. The operator $\otimes$ denotes the Kronecker product, $\mathbf{s}_k \in \mathbb{C}^{L\times 1}$ is the unit-norm spreading signature, and $x_k$ is the transmitted symbol. The term $\mathbf{z} \in \mathbb{C}^{N_rL \times 1}$ is the sum of all received signals from outside the intended cluster, i.e., intercluster interference, and $\mathbf{n} \in \mathbb{C}^{N_rL \times 1}$ is the zero-mean complex Gaussian noise with variance $\sigma^2_{\mathbf{n}}$. The description using the Kronecker product follows from the fact that at each \ac{BS} antenna, a spreading block (e.g., over neighbouring subcarriers) of length $L$ is received, and therefore the total received signal (over space and frequency) is of size $N_rL$. This, combined with the assumption of the channel being flat over the spreading block, allows for this compact description. Since the \ac{RIS} is deployed in a \ac{LOS} to the \ac{BS}, the \ac{BS}-\ac{RIS} channel is rank-1, and is given by $\mathbf{H}_{\textrm{BS}} = \mathbf{a}\mathbf{b}^H$, where $\mathbf{a}$ and $\mathbf{b}$ are the array responses at the \ac{BS} and \ac{RIS}, respectively. The received signal can then be written as
\begin{align}\label{eq:03}
	\mathbf{y} = (\mathbf{a} \otimes \mathbf{I}_L) \sum_{k = 1}^{K} \sqrt{\ell_{\textrm{BS}} \ell_{h_k} P_k L}\, \big(\mathbf{b}^H\mathbf{\Phi}\mathbf{h}_{k} \otimes \mathbf{s}_k\big)x_k + \mathbf{z} + \mathbf{n},
\end{align}
where $\mathbf{I}_L$ is the identity matrix of size $L$. Equipped with a large number of antennas, the \ac{BS} forms a beam towards the cluster's \ac{RIS}, boosting the received power on the one hand, and on the other hand, suppressing intercluster interference. To achieve that, beam forming via \ac{MRC} is performed. This further justifies the assumption of dropping the direct path between the \acp{UE} and the \ac{BS}, as it would be even weaker after beamforming. The \ac{MRC} spatially filtered signal $\tilde{\mathbf{y}} = (\mathbf{a}^H \otimes \mathbf{I}_L) \mathbf{y}$ is given by
\begin{align}\label{eq:04}
	\begin{split}
		\tilde{\mathbf{y}} =\, &(\mathbf{a}^H\mathbf{a} \otimes \mathbf{I}_L) \sum_{k = 1}^{K} \sqrt{\ell_{\textrm{BS}} \ell_{h_k} P_k L}\, \big(\mathbf{b}^H\mathbf{\Phi}\mathbf{h}_{k} \otimes \mathbf{s}_k\big)x_k \\ &+ (\mathbf{a}^H \otimes \mathbf{I}_L)\mathbf{z} + (\mathbf{a}^H \otimes \mathbf{I}_L)\mathbf{n}.
	\end{split}
\end{align}
With the beamforming towards the target \ac{RIS}, intercluster interference is greatly reduced, i.e., $(\mathbf{a}^H \otimes \mathbf{I}_L)\mathbf{z} \approx \mathbf{0}$. Since we have $\mathbf{a}^H\mathbf{a} = N_r$, and letting $\tilde{\mathbf{n}} = (\mathbf{a}^H \otimes \mathbf{I}_L)\mathbf{n}$ be the spatially filtered noise with $\sigma^2_{\tilde{\mathbf{n}}} = N_r\sigma^2_{\mathbf{n}}$, \eqref{eq:04} is further developed as
\begin{align}\label{eq:05}
	\tilde{\mathbf{y}} = \sum_{k = 1}^{K} \sqrt{N_r^2\ell_{\textrm{BS}} \ell_{h_k} P_k L}\, \big(\mathbf{b}^H\mathbf{\Phi}\mathbf{h}_{k} \otimes \mathbf{s}_k\big)x_k + \tilde{\mathbf{n}}.
\end{align}
Notice that $\mathbf{b}^H\mathbf{\Phi}\mathbf{h}_{k}$ is a scalar and therefore $\otimes$ is no longer necessary. Let $\beta_k = \sqrt{N_r^2\ell_{\textrm{BS}} \ell_{h_k} P_k L}$, $\mathbf{w} = \textrm{diag}(\mathbf{\Phi}^H)$, and $\hat{\mathbf{h}}_{k} = \mathbf{b}^* \circ \mathbf{h}_{k}$, where $\circ$ denotes the Hadamard product, the post-spatially filtered signal can finally be written as
\begin{align}\label{eq:06}
	\tilde{\mathbf{y}} = \sum_{k = 1}^{K} \beta_k (\mathbf{w}^H \hat{\mathbf{h}}_{k})\,\mathbf{s}_kx_k + \tilde{\mathbf{n}}.
\end{align}
In order to detect the \acp{UE} within the cluster, the \ac{BS} performs \ac{MMSE}-\ac{IC} detection, with a \ac{UE} being detected correctly if its \ac{SINR} exceeds a certain rate threshold. Assuming a successive \ac{IC} in which one \ac{UE} is detected per \ac{IC} stage, and assuming a detection order of $\text{UE }1,\,\text{UE }2,\,\dots,\,\text{UE }K$, the post-filtering \ac{SINR} of the $k^{\text{th}}$ \ac{UE} is given by
\begin{align}\label{eq:07}
	\textrm{SINR}_k = \frac{|\beta_k \big(\mathbf{w}^H \hat{\mathbf{h}}_k\big) \mathbf{v}_k^H \mathbf{s}_k^{}|^2}{\sum_{l = k + 1}^{K} |\beta_l \big(\mathbf{w}^H \hat{\mathbf{h}}_l\big) \mathbf{v}_k^H \mathbf{s}_l^{}|^2 + \sigma^2_{\tilde{\mathbf{n}}}\lVert\mathbf{v}_k\lVert^2},
\end{align}
where $\mathbf{v}_k$ is the \ac{MMSE} filter applied at the $k^{\text{th}}$ stage. As can be seen, every time a \ac{UE} is removed, the next \ac{UE} in the next \ac{IC} stage experiences less interference, until we reach the last \ac{UE}, in which it only has to deal with noise. The goal now is to design $\mathbf{w}$ such that 
\begin{align}\label{eq:08}
	\textrm{SINR}_k \ge \epsilon_k, \quad \forall k,~k = 1,\,2,\,\dots,\,K,
\end{align}
where $\epsilon_k$ is the detection threshold of the $k^{\textrm{th}}$ \ac{UE}. In other words, we choose the phase-shifts at the \ac{RIS} such that the power gaps between the \acp{UE} combined with the \ac{MMSE} filtering and \ac{IC} result in \acp{SINR} exceeding the required threshold for decodability, at each of the \ac{IC} stages. 

We have multiple problems here; first, the detection order of the \acp{UE}, to begin with, is unknown and it depends on the choice of $\mathbf{w}$. This can be clearly seen in \eqref{eq:06}, where the received power of the users is directly impacted by the choice of $\mathbf{w}$. In other words, the optimal detection order and $\mathbf{w}$ need to be determined jointly, requiring a search over all possible detection orders, which can be of prohibitive complexity for large $K$; second, the resulting \ac{SINR} at each stage depends on the \ac{MMSE} filter $\mathbf{v}_k$; however, $\mathbf{v}_k$ also depends on $\mathbf{w}$ and the detection order (coupled), and therefore determining $\mathbf{w}$ depends on the resulting $\mathbf{v}_k$; third, even if everything is known, how do we find a $\mathbf{w}$ satisfying all of the $K$ inequalities in \eqref{eq:08}? 

\section{Sum-Rate Optimized Phase-Shifts}\label{sec:03}
It is known from the \ac{MIMO} literature that \ac{MMSE}-\ac{IC} is a sum-rate optimal detection scheme \cite{Tse05}. Therefore, one way to avoid the aforementioned problems with the detection order and the choice of the \ac{MMSE} filter, is to optimize $\mathbf{w}$ such that the sum-rate of the cluster is maximized. To that end, the sum-rate is given by
\begin{align}\label{eq:09}
	R_{\textrm{sum}} = \frac{1}{L}\log_2 \textrm{det}\bigg(\mathbf{I}_L + \frac{1}{\sigma^2_{\tilde{\mathbf{n}}}} \sum_{k = 1}^{K} \beta_k^2(\mathbf{w}^H \hat{\mathbf{h}}_{k}^{}) \mathbf{s}_{k}^{} \mathbf{s}_{k}^H (\hat{\mathbf{h}}_{k}^H\mathbf{w})   \bigg).
\end{align}
Due to the determinant operator $\textrm{det}()$ and the $\mathbf{s}_{k}^{} \mathbf{s}_{k}^H$ term, maximizing the above sum-rate expression is not an easy task. To manage that, we drop the spreading, and optimize the system as if no spreading is employed, i.e., we set $L = 1$ and $\mathbf{s}_k = 1, \forall k$. Such an optimization would correspond to a pure power-domain \ac{NOMA} system, i.e., a worst-case scenario in which the spreading has no impact. Then, \eqref{eq:09} becomes
\begin{align}\label{eq:10}
	R_{\textrm{sum}}^{(\text{no spread.})} = \log_2 \bigg(1 + \frac{1}{\sigma^2_{\tilde{\mathbf{n}}}} \sum_{k = 1}^{K} \beta_k^2\mathbf{w}^H \hat{\mathbf{h}}_{k}^{} \hat{\mathbf{h}}_{k}^H\mathbf{w}   \bigg).
\end{align}
Let $\mathbf{H} = \sum_{k = 1}^{K} \beta_k^2 \hat{\mathbf{h}}_{k}^{} \hat{\mathbf{h}}_{k}^H$, the sum-rate maximizer is given by
\begin{align}\label{eq:11}
	\begin{split}
		\mathbf{w}_{\textrm{sum}} =&\argmax_{\mathbf{w}} ~ \mathbf{w}^H \mathbf{H} \mathbf{w} \\
		&\,\text{s.t.}~~|[\mathbf{w}]_n| = 1, \quad n = 1,2, \dots, N_s,
	\end{split}	
\end{align}
where the condition $|[\mathbf{w}]_n| = 1$ refers to the $n^{\textrm{th}}$ element of $\mathbf{w}$ performing a phase-shift only. In order to solve \eqref{eq:11}, we relax it to a conventional quadratic problem. Therefore, the maximizer of $\mathbf{w}^H \mathbf{H} \mathbf{w}$ is given by the eigenvector of $\mathbf{H}$ corresponding to its maximum eigenvalue. Let $\mathbf{u}_{\textrm{max}}$ be that eigenvector, the elements of $\mathbf{w}_{\textrm{sum}}$ are then set to
\begin{align}\label{eq:12}
	[\mathbf{w}_{\textrm{sum}}]_n = \exp(j \angle [\mathbf{u}_{\textrm{max}}]_n), \quad n = 1,2, \dots, N_s,
\end{align}
i.e., $\mathbf{w}_{\textrm{sum}}$ is set such that it performs the same phase-shifts as $\mathbf{u}_{\textrm{max}}$. The issue with the sum-rate optimized shifts is that if the \acp{UE} have similar receive powers, then the \ac{RIS} would boost all of them by an equal amount, i.e., it only provides a \ac{SNR} gain (the strongest eigenvector would point in the direction that favors all the \acp{UE}). This is beneficial if the system suffers from low \ac{SNR}; however, our major problem here is multi-user interference, and the goal is to boost the \acp{UE} with different portions, such that sufficient power gaps are created between them, allowing the \ac{IC} to operate successfully. Also, in our optimization above, spreading is not taken into account. However, if the \acp{UE} have sufficient power gaps between them (e.g., due to different pathlosses), then $\mathbf{w}_{\textrm{sum}}$ can provide a good solution, as the strongest eigenvector would point in the direction of the strongest \acp{UE}, and this helps to further enlarge the gaps (the \ac{RIS} would boost the stronger \acp{UE} further), resulting in better sequential \acp{SINR} under IC. We will see this effect later in \Cref{sec:04}.
\vspace{-0mm}
\section{Proposed Optimization Approach}
Robust optimization of the phase-shifts requires solving the inequalities of \eqref{eq:08}. However, as we mentioned before, the optimal solution is difficult to obtain, due to the coupling between the detection order and the \ac{MMSE} filter with our $\mathbf{w}$. In the following, we propose a suboptimal procedure that allows us to obtain a solution to the problem.
\vspace{-2mm}
\subsection{Detection Order}
The optimal solution requires an exhaustive search over all possible detection orders, consisting of $K!$ possibilities. This can be prohibitive for large $K$, and it is the large $K$ that we are interested in. A suboptimal approach that can provide a good performance \cite{Yang20}, is to order the \acp{UE} based on their received signal strength, i.e, $|\beta_k \mathbf{w}^H \hat{\mathbf{h}}_{k}|$. However, we can see that it depends on $\mathbf{w}$, which we seek to find in the first place. For that reason, we do the ordering based on the sum-rate optimized shifts, i.e., by ordering the \acp{UE} according to $|\beta_k \mathbf{w}_{\textrm{sum}}^H \hat{\mathbf{h}}_{k}|$. In other words, $\mathbf{w}_{\textrm{sum}}$ is employed as the initial solution for determining the detection order. In the following, and without loss of generality, we assume the resultant \acp{UE} ordering is
\begin{align}\label{eq:13}
	|\beta_1 \mathbf{w}_{\textrm{sum}}^H \hat{\mathbf{h}}_{1}| \ge |\beta_2 \mathbf{w}_{\textrm{sum}}^H \hat{\mathbf{h}}_{2}| \ge \dots \ge |\beta_K \mathbf{w}_{\textrm{sum}}^H \hat{\mathbf{h}}_{K}|,
\end{align}
that is, after ordering, $\text{UE }1$ is the strongest user, while $\text{UE }K$ is the weakest one. This assumption is only applied to simplify notation for the next parts.
\subsection{\ac{MMSE} Filtering}
The next coupled variable is the \ac{MMSE} filter. We follow a similar approach as with the detection order. We calculate the \ac{MMSE} filters based on the sum-rate solution. Therefore, given our determined detection order and $\mathbf{w}_{\textrm{sum}}$, the \ac{MMSE} filters applied in \eqref{eq:07} are such that
\begin{align}\label{eq:14}
	\begin{split}
		\mathbf{v}_k^H = \mathbf{g}^H_k\bigg(\sum_{l = k}^{K} \mathbf{g}_l^{} \mathbf{g}^H_l + \mathbf{I}_L \sigma^2_{\tilde{\mathbf{n}}}\bigg)^{-1},
	\end{split}
\end{align}
where $\mathbf{g}_k = \beta_k (\mathbf{w}_{\textrm{sum}}^H \hat{\mathbf{h}}_{k})\,\mathbf{s}_k$.
\subsection{Phase-Shifts Optimization}
Having both the detection order and \ac{MMSE} filter determined based on $\mathbf{w}_{\textrm{sum}}$, we now proceed to finding our final phase-shifts. First, we rewrite \eqref{eq:07} as
\begin{align}\label{eq:15}
	\frac{\mathbf{w}^H \Big(\beta^2_k |\mathbf{v}_k^H \mathbf{s}_k^{}|^2\, \hat{\mathbf{h}}_k^{} \hat{\mathbf{h}}_k^H\Big) \mathbf{w}}{\mathbf{w}^H \Big( \sum_{l = k + 1}^{K} \beta^2_l |\mathbf{v}_k^H \mathbf{s}_l^{}|^2\, \hat{\mathbf{h}}_l^{} \hat{\mathbf{h}}_l^H + \frac{\sigma^2_{\tilde{\mathbf{n}}}\lVert\mathbf{v}_k\lVert^2}{N_s}\mathbf{I}_{N_s} \Big)\mathbf{w}},
\end{align}
where the fact that $\mathbf{w}^H\mathbf{w} = N_s$ has been applied to the noise term. Let
\begin{align}\label{eq:16}
	\begin{split}
		\mathbf{A}_k &= \beta^2_k |\mathbf{v}_k^H \mathbf{s}_k^{}|^2\, \hat{\mathbf{h}}_k^{} \hat{\mathbf{h}}_k^H,\\
		\mathbf{B}_k &= \sum_{l = k + 1}^{K} \beta^2_l |\mathbf{v}_k^H \mathbf{s}_l^{}|^2\, \hat{\mathbf{h}}_l^{} \hat{\mathbf{h}}_l^H + \frac{\sigma^2_{\tilde{\mathbf{n}}}\lVert\mathbf{v}_k\lVert^2}{N_s}\mathbf{I}_{N_s},
	\end{split}
\end{align}
our optimization problem is then formulated as 
\begin{align}\label{eq:17}
	\begin{split}
		\text{find}~~&\mathbf{w} \\
		\text{s.t.}~~&\frac{\mathbf{w}^H \mathbf{A}_k \mathbf{w}}{\mathbf{w}^H \mathbf{B}_k\mathbf{w}}  \ge \epsilon_k, \quad k = 1,\,2,\,\dots,\,K,\\
		& |[\mathbf{w}]_n| = 1, \quad \quad~~~n = 1,\,2,\,\dots,\,N_s.
	\end{split}	
\end{align}
To find a solution to those series of inequalities, we relax \eqref{eq:17} into a \ac{SDP} problem, which can be solved efficiently using convex optimization algorithms \cite{Luo10}. Let $\mathbf{W} = \mathbf{w}\mathbf{w}^H$; using the trace operator, we have $\mathbf{w}^H \mathbf{A}_k \mathbf{w} = \textrm{tr}\big(\mathbf{A}_k \mathbf{w}\mathbf{w}^H\big) = \textrm{tr}\big(\mathbf{A}_k \mathbf{W}\big)$. Similarly, we have $\mathbf{w}^H \mathbf{B}_k\mathbf{w} = \textrm{tr}\big(\mathbf{B}_k \mathbf{W}\big)$. The \ac{SINR} condition is then written as
\begin{align}
	\begin{split}
		\textrm{tr}\big(\mathbf{A}_k \mathbf{W}\big) - \epsilon_k\, \textrm{tr}\big(\mathbf{B}_k \mathbf{W}\big) &\ge 0, \\
		\textrm{tr}\Big(\big[\mathbf{A}_k - \epsilon_k\mathbf{B}_k \big]\mathbf{W}\Big) &\ge 0.
	\end{split}
\end{align}
Finally, our \ac{SDP}-relaxed problem is given by
\begin{align}\label{eq:19}
	\begin{split}
		\text{find}~~&\mathbf{W} \\
		\text{s.t.}~~&\textrm{tr}\Big(\big[\mathbf{A}_k - \epsilon_k\mathbf{B}_k \big]\mathbf{W}\Big) \ge 0, \quad k = 1,\,2,\,\dots,\,K,\\
		&\mathbf{W} \succcurlyeq 0,\,[\mathbf{W}]_{n,n} = 1, \quad \quad~~\, n = 1,\,2,\,\dots,\,N_s,
	\end{split}
\end{align}
where $[\mathbf{W}]_{n,n}$ is the $n^{\textrm{th}}$ diagonal element of $\mathbf{W}$. In this work, the optimizer used is based on CVX \cite{cvx}. If a solution is found, then we set $\mathbf{w}_{\textrm{prop}}$ (proposed) such that it performs the same phase-shifts as the eigenvector of $\mathbf{W}$ corresponding to its maximum eigenvalue (best rank-1 approximation) in a similar fashion as in \eqref{eq:12}. If no solution is feasible, then we rely on the sum-rate solution, i.e., we set $\mathbf{w}_{\textrm{prop}} = \mathbf{w}_{\textrm{sum}}$.

\section{Investigation of an Example Scenario}\label{sec:04}
We consider a scenario where $K$ active \acp{UE} in the target cluster communicate with a $32$-antennas \ac{BS} through a $32$-elements \ac{RIS}. A $4 \times 16$ Grassmannian codebook is employed for the spreading, i.e, the signature length is $L = 4$ with a total of $16$ signatures in the codebook. The Grassmannian criterion refers to designing the codebook such that the maximum cross-correlation between any pair of signatures is minimized \cite{Tahir19b}. The \ac{BS}-\ac{RIS} channel is \ac{LOS} with pathloss $\ell_{\mathrm{BS}} = -65$\,dB, while the \ac{RIS}-\ac{UE} channels are modeled as Rayleigh fading with the pathloss uniformly disturbed as $\ell_{h_k} \sim \mathcal{U}(-65-s, -65+s)$\,dB, i.e., a mean component of $-65$\,dB plus a spread of $\pm s$. By adjusting $s$, we can control the pathloss differences across the \acp{UE}, and thus the average received power difference between the \acp{UE} at the \ac{BS}. We assume the \acp{UE} to transmit with an equal power of $P_k = P = 30$\,dBm, $\forall k$. The noise power is set to $\sigma^2_{\mathbf{n}} = -110$\,dBm. Also, we assume all the \acp{UE} to have the same threshold $\epsilon_k = \epsilon$. The simulation points are given with $95\%$ confidence intervals shown as gray bars.

In \Cref{fig:01}, we compare the detection performance using our proposed approach versus the sum-rate optimized phase-shifts and random ones. The results are shown for a pathloss spread of $\pm 3$\,dB, and over the outage thresholds of $\epsilon = 1,\,4,\,$and $9$\,dB. The desired result here is a $1$:$1$ line, i.e., all active \acp{UE} are detected correctly. We observe that our proposed \ac{RIS} adaptation allows for a substantial increase of the number of correctly detected users. This is achieved at both low and high outage thresholds. As the threshold increases, it becomes more challenging for the \ac{RIS} to satisfy all the inequalities of \eqref{eq:08}. If the threshold is too high for the number of active \acp{UE}, then no feasible solution would be possible, and the number of correctly detected \acp{UE} begins to drop.

\begin{figure}[t]
	\centering
	\resizebox{0.93\linewidth}{!}{%
		\pgfplotsset{width=250pt, height=200pt, compat = 1.9}
		\begin{tikzpicture}
	\begin{axis}[
	scale only axis,
	unit vector ratio*=1 1 1,
	xlabel={Active UEs $K$},
	ylabel={Correctly deteceted UEs},
	label style={font=\labelfont},
	ylabel shift = -1mm,
	ylabel right ={~},
	ymin=0, ymax=16,
	xmin=0, xmax=16,
	xtick={0,2,4,6,8,10,12,14,16},
	ytick={0,2,4,6,8,10,12,14,16},
	ytick pos=left,
	axisSetup,
	ymajorgrids=true,
	xmajorgrids=true,
	yminorgrids=true,
	xminorgrids=true,
	major x grid style={solid, cgrid},
	major y grid style={solid, cgrid},
	minor x grid style={solid, cgrid},
	minor y grid style={gridc},
	legend style={font=\legendfont, name=legendNode},
	legend cell align=left,
	legend pos=north west,
	]
	\foreach \i/\c/\m in {1/linew/*, 2/linew3/*, 3/linew2/*}{
		\edef\temp{\noexpand \addplot [\c, color=blueC1, mark=*, mark options={errorlinemarkSty}, forget plot] [error bars/.cd, y dir=both, y explicit, error mark options={errormarkSty}, error bar style={errorbarSty}] table [x=x\i, y=y\i, y error minus=cl\i, y error plus=ch\i, col sep=comma] {graphics/results/fig_UEdetect.csv};}
		\temp
	};
	\foreach \i/\c/\m in {4/linew/*, 5/linew3/*, 6/linew2/*}{
		\edef\temp{\noexpand \addplot [\c, color=redC1, mark=*, mark options={errorlinemarkSty}, forget plot] [error bars/.cd, y dir=both, y explicit, error mark options={errormarkSty}, error bar style={errorbarSty}] table [x=x\i, y=y\i, y error minus=cl\i, y error plus=ch\i, col sep=comma] {graphics/results/fig_UEdetect.csv};}
		\temp
	};
	\foreach \i/\c/\m in {7/linew/squ*, 8/linew3/*, 9/linew2/*}{
		\edef\temp{\noexpand \addplot [\c, color=greenC1, mark=*, mark options={errorlinemarkSty}, forget plot] [error bars/.cd, y dir=both, y explicit, error mark options={errormarkSty}, error bar style={errorbarSty}] table [x=x\i, y=y\i, y error minus=cl\i, y error plus=ch\i, col sep=comma] {graphics/results/fig_UEdetect.csv};}
		\temp
	};
	\addlegendimage{short Legend0, solid};
	\addlegendentry{Proposed \hspace*{-4pt}};
	\addlegendimage{short Legend1, dash pattern=on 1pt off 2pt};
	\addlegendentry{Sum-rate \hspace*{-4pt}};
	\addlegendimage{short Legend0, dash pattern=on 5pt off 4pt};
	\addlegendentry{Random \hspace*{-4pt}};
	%
	\draw (rel axis cs: 0.64, 0.875) node[] {{\legendfont \contourlength{1.5pt} \contour*{white}{$\epsilon = 1$\,dB}}};
	\draw [-{Circle[scale=0.75]}, line width=0.75pt, color=dark01](rel axis cs: 0.74, 0.875) -- (rel axis cs: 0.915, 0.91);
	\draw [-{Circle[scale=0.75]}, line width=0.75pt, color=dark01](rel axis cs: 0.74, 0.865) -- (rel axis cs: 0.91, 0.852);
	\draw [-{Circle[scale=0.75]}, line width=0.75pt, color=dark01](rel axis cs: 0.72, 0.855) -- (rel axis cs: 0.835, 0.688);
	\draw (rel axis cs: 0.877, 0.435) node[] {{\legendfont \contourlength{1.5pt} \contour*{white}{$\epsilon = 4$\,dB}}};
	\draw [-{Circle[scale=0.75]}, line width=0.75pt, color=dark01](rel axis cs: 0.9, 0.465) -- (rel axis cs: 0.955, 0.78);
	\draw [-{Circle[scale=0.75]}, line width=0.75pt, color=dark01](rel axis cs: 0.86, 0.46) -- (rel axis cs: 0.71, 0.52);
	\draw [-{Circle[scale=0.75]}, line width=0.75pt, color=dark01](rel axis cs: 0.85, 0.41) -- (rel axis cs: 0.705, 0.24);
	\draw (rel axis cs: 0.26, 0.5) node[] {{\legendfont \contourlength{1.5pt} \contour*{white}{$\epsilon = 9$\,dB}}};
	\draw [-{Circle[scale=0.75]}, line width=0.75pt, color=dark01](rel axis cs:  0.36, 0.5) -- (rel axis cs: 0.535, 0.49);
	\draw [-{Circle[scale=0.75]}, line width=0.75pt, color=dark01](rel axis cs:  0.35, 0.48) -- (rel axis cs: 0.42, 0.35);
	\draw [-{Circle[scale=0.75]}, line width=0.75pt, color=dark01](rel axis cs:  0.32, 0.475) -< (rel axis cs: 0.37, 0.3);
	\end{axis}
\end{tikzpicture}
	}
	\caption{Detectability under different strategies. Here $N_s = 32$, $P = 30\,$dBm, and pathloss spread is $\pm 3$\,dB.}
	\label{fig:01}
	\bigbreak
	\centering
	\resizebox{0.93\linewidth}{!}{%
		\pgfplotsset{width=250pt, height=210pt, compat = 1.9}
		\begin{tikzpicture}
	\begin{axis}[
	xlabel={Pathloss spread $s$ [dB]},
	ylabel={Correctly deteceted UEs},
	label style={font=\labelfont},
	ylabel shift = -1mm,
	ylabel right ={~},
	ymin=0, ymax=12,
	xmin=0, xmax=6,
	xtick={0,1,2,3,4,5,6},
	ytick={0,2,4,6,8,10,12},
	ytick pos=left,
	axisSetup,
	ymajorgrids=true,
	xmajorgrids=true,
	yminorgrids=true,
	xminorgrids=true,
	major x grid style={solid, cgrid},
	major y grid style={solid, cgrid},
	minor x grid style={solid, cgrid},
	minor y grid style={gridc},
	legend style={font=\legendfont, name=legendNode, at={(0.02,0.5)},anchor=south west},
	legend cell align=left,
	]
	\foreach \i/\c/\m in {1/linew/o, 2/linew3/o, 3/linew2/o}{
		\edef\temp{\noexpand \addplot [\c, color=blueC1, mark=*, mark options={errorlinemarkSty}, forget plot] [error bars/.cd, y dir=both, y explicit, error mark options={errormarkSty}, error bar style={errorbarSty}] table [x=x\i, y=y\i, y error minus=cl\i, y error plus=ch\i, col sep=comma] {graphics/results/fig_pathlossSpread.csv};}
		\temp
	};
	\foreach \i/\c/\m in {4/linew/square, 5/linew3/square, 6/linew2/square}{
		\edef\temp{\noexpand \addplot [\c, color=redC1, mark=*, mark options={errorlinemarkSty}, forget plot] [error bars/.cd, y dir=both, y explicit, error mark options={errormarkSty}, error bar style={errorbarSty}] table [x=x\i, y=y\i, y error minus=cl\i, y error plus=ch\i, col sep=comma] {graphics/results/fig_pathlossSpread.csv};}
		\temp
	};
	\addlegendimage{short Legend0, solid};
	\addlegendentry{Proposed \hspace*{-4pt}};
	\addlegendimage{short Legend1, dash pattern=on 1pt off 2pt};
	\addlegendentry{Sum-rate \hspace*{-4pt}};
	\addlegendimage{short Legend0, dash pattern=on 5pt off 4pt};
	\addlegendentry{Random \hspace*{-4pt}};
	\draw (rel axis cs: 0.85, 0.6) node[] {{\legendfont \contourlength{1.5pt} \contour*{white}{\begin{tabular}{c} NOMA \\ codebook \end{tabular}}}};
	\draw [-{Circle[scale=0.75]}, line width=0.75pt, color=dark01](rel axis cs: 0.835, 0.68) -- (rel axis cs: 0.75, 0.99);
	\draw [-{Circle[scale=0.75]}, line width=0.75pt, color=dark01](rel axis cs: 0.87, 0.68) -- (rel axis cs: 0.9, 0.85);
	\draw [-{Circle[scale=0.75]}, line width=0.75pt, color=dark01](rel axis cs: 0.865, 0.55) -- (rel axis cs: 0.915, 0.39);
	\draw (rel axis cs: 0.5, 0.47) node[] {{\legendfont \contourlength{1.5pt} \contour*{white}{\begin{tabular}{c} OMA \\ codebook \end{tabular}}}};
	\draw [-{Circle[scale=0.75]}, line width=0.75pt, color=dark01](rel axis cs: 0.5, 0.55) -- (rel axis cs: 0.45, 0.81);
	\draw [-{Circle[scale=0.75]}, line width=0.75pt, color=dark01](rel axis cs: 0.6, 0.45) -- (rel axis cs: 0.7, 0.40);
	\draw [-{Circle[scale=0.75]}, line width=0.75pt, color=dark01](rel axis cs: 0.54, 0.41) -- (rel axis cs: 0.63, 0.212);
	\end{axis}
\end{tikzpicture}
	}
	\caption{Impact of the pathloss spread and codebook design on the performance. Here $N_s = 32$, $K = 12$, and $\epsilon = 4$\,dB.}
	\label{fig:02}
	\vspace{-4mm}
\end{figure}
\begin{figure}[t]
	\centering
	\resizebox{0.92\linewidth}{!}{%
		\pgfplotsset{width=250pt, height=210pt, compat = 1.9}
		\begin{tikzpicture}
\begin{semilogxaxis}[
xlabel={Number of elements $N_s$},
ylabel={Correctly deteceted UEs},
label style={font=\labelfont},
ylabel shift = -1mm,
ylabel right ={~},
ymin=0, ymax=12,
xmin=0, xmax=100,
ytick={0,2,4,6,8,10,12},
ytick pos=left,
axisSetup,
ymajorgrids=true,
xmajorgrids=true,
yminorgrids=true,
xminorgrids=true,
major x grid style={solid, cgrid},
major y grid style={solid, cgrid},
minor x grid style={dotted},
minor y grid style={gridc},
legend style={font=\legendfont, name=legendNode},
legend cell align=left,
legend pos=north west,
]
\foreach \i/\c/\m in {1/linew/o, 2/linew3/o, 3/linew2/o}{
	\edef\temp{\noexpand \addplot [\c, color=blueC1, mark=*, mark options={errorlinemarkSty}, forget plot] [error bars/.cd, y dir=both, y explicit, error mark options={errormarkSty}, error bar style={errorbarSty}] table [x=x\i, y=y\i, y error minus=cl\i, y error plus=ch\i, col sep=comma] {graphics/results/fig_nElements.csv};}
	\temp
};
\foreach \i/\c/\m in {4/linew/square, 5/linew3/square, 6/linew2/square}{
	\edef\temp{\noexpand \addplot [\c, color=redC1, mark=*, mark options={errorlinemarkSty}, forget plot] [error bars/.cd, y dir=both, y explicit, error mark options={errormarkSty}, error bar style={errorbarSty}] table [x=x\i, y=y\i, y error minus=cl\i, y error plus=ch\i, col sep=comma] {graphics/results/fig_nElements.csv};}
	\temp
};
\addlegendimage{short Legend0, solid};
\addlegendentry{Proposed \hspace*{-4pt}};
\addlegendimage{short Legend1, dash pattern=on 1pt off 2pt};
\addlegendentry{Sum-rate \hspace*{-4pt}};
\addlegendimage{short Legend0, dash pattern=on 5pt off 4pt};
\addlegendentry{Random \hspace*{-4pt}};
%
\draw (rel axis cs: 0.59, 0.072) [rotate=45] arc(-120:120:4pt and 20pt) [line width=0.7pt] [rotate=-45] node[pos= 0.45, right=-75pt] {{\legendfont \contourlength{1.5pt} \contour*{white}{$P = -5$\,dBm}}};
\draw (rel axis cs: 0.28, 0.34) [rotate=10] arc(-120:120:4pt and 22pt) [line width=0.7pt] [rotate=-10] node[pos= 0.5, right=-20pt] {{\legendfont \contourlength{1.5pt} \contour*{white}{$P = 30$\,dBm}}};
\end{semilogxaxis}
\end{tikzpicture}
	}
	\caption{Scaling of the performance with $N_s$. Here $K = 12$, $\epsilon = 3$\,dB, and pathloss spread is $\pm 0$\,dB.}
	\label{fig:03}
	\vspace{-6mm}
\end{figure}
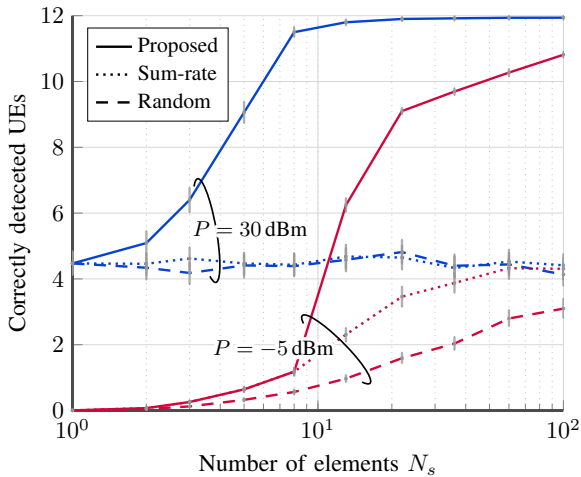

Next, we set $K = 12$ and $\epsilon = 4$\,dB and investigate the performance over different pathloss spreads. We also compare our \ac{NOMA} (Grassmannian) codebook to an \ac{OMA} codebook, e.g., a $4\times 4$ identity matrix. In the case of the \ac{OMA} codebook, we only have $4$ signatures, and therefore unique signature assignment to the $12$ \acp{UE} is not possible, i.e., the orthogonal signatures must be reused between the users. The results are shown in \Cref{fig:02}. We observe that, at least for the considered configuration, the \ac{NOMA} codebook offers substantial improvement over the \ac{OMA} codebook reuse strategy, across the different adaptation approaches. The \ac{NOMA} codebook distributes the interference across all the signatures, which leads to improved performance under \ac{IC}, as canceling one \ac{UE} would help in improving the \ac{SINR} of all remaining \acp{UE}, and not just to a limited subset as would occur when reusing the \ac{OMA} codebook. We see that our approach provides a robust adaptation of the \ac{RIS} phase-shifts with respect to the pathloss spread, and is able to create the necessary power gaps that result in the required \ac{SINR} levels. As for the sum-rate optimized phases, we observe that the performance improves as the pathloss spread increases. As explained in \Cref{sec:03}, the power gap between the \acp{UE} resulting from the larger pathloss spreads goes in the favor of the sum-rate solution. At low pathloss spreads, their user-separability performance approaches that of the random shifts. The gain at those ranges is mostly an \ac{SNR} gain, which is not visible in the figure due to the relatively high transmit power.

To further investigate that, in \Cref{fig:03} we show the performance over the number of \ac{RIS} elements $N_s$, for low and high transmit powers of $P = $ $-5\,$dBm and $30\,$dBm, respectively. We set the pathloss spread to $\pm 0$\,dB and $\epsilon = 3\,$dB. First, we make the observation that a certain number of elements is required in order for \eqref{eq:08} to be solvable; second, at low transmit powers and $0$\,dB pathloss spread, the sum-rate optimized phase-shifts clearly provide an \ac{SNR} gain compared to the random phase-shifts, converging towards the performance of that at high transmit power as $N_s$ increases. Our approach, as can be seen, is capable of providing both \ac{SNR} and \ac{SINR} gains under \ac{IC}.

\section{Discussion}
\vspace{-2mm}
We discuss in this section some aspects regarding our approach and assumptions, and possible further improvements. 
\begin{itemize}[leftmargin=0pt]
	\item[] -- The spatial filtering applied is based on \ac{MRC}, following the assumption of the \ac{BS} being equipped with a large antenna array. If strong intercluster interference is present, then \ac{MMSE}-based spatial filtering is a possibility. The model in \eqref{eq:06} would still hold, except that an \ac{MMSE} filter is applied instead of $\mathbf{a}^H$. Another solution, which is based on the code-domain, is to jointly design the spreading codebooks across the different clusters \cite{Tahir19a}, such that the cross-correlation between the signatures of the neighbouring clusters is reduced. 
	
	\item[] -- Throughout our optimization approach, we assumed the \acp{UE} to have fixed transmit powers. Power control can help improve the performance of the uplink, by adjusting the transmit power of \acp{UE} depending on their channel conditions. Therefore, further improvement of the performance can be achieved by jointly optimizing the \ac{RIS} phase-shifts and the transmit powers.
	\item[] -- The assignment of the signatures to the \acp{UE} can also have an impact on the performance. In this work, we assumed a fixed signature assignment; still, we can utilize the structure of non-uniform codebooks such that \acp{UE} close in receive power are assigned near-orthogonal signatures, which then allows to suppress the interference between them by filtering. On the other hand, \acp{UE} with large power gaps can be assigned the same signature, since those can be separated via \ac{IC}. However, this is also coupled with our phase-shifts optimization.
	\item[] -- Although we assumed the use of successive \ac{MMSE}-\ac{IC} detection, several alternatives are possible. For example, parallel \ac{IC} can be performed, allowing multiple \acp{UE} to be detected per \ac{IC} iteration, which reduces the detection latency for large $K$. Also, under certain conditions, the \ac{NOMA} filtering can be performed in a low-complexity manner \cite{Tahir20a}.
	\item[] -- For the extraction of the final phase-shifts in \eqref{eq:19}, a rank-1 approximation based on the strongest eigenvector was applied. In general, better approximation might be achieved via a Gaussian randomization procedure \cite{Luo10}.
\end{itemize} 

\section{Conclusion}

We investigate the combination of code-domain \ac{NOMA} and \acp{RIS} in a massive \ac{MIMO} cluster-based deployment.  In order to support multi-user transmission within the clusters, code-domain \ac{NOMA} is employed. We consider the optimization of the \ac{RIS} such that a large number of users are detected correctly. To overcome the coupling between the \ac{RIS} phase-shifts and other variables, such as the detection order and the applied filters, we utilize sum-rate optimized phase-shifts to obtain a decoupled estimate of those variables. Given the now-decoupled problem, the final phase-shifts are found by an \ac{SDP} relaxation of the problem, which can be solved efficiently via convex optimization methods. Simulations results show that our approach is an effective \ac{RIS} adaptation strategy.

\begin{acronym}[DSTTDSGRC]
\setlength{\itemsep}{-3pt}
\acro{CS}{compressed sensing}
\acro{ETF}{equiangular tight frame}
\acro{OGF}{orthoplectic Grassmannian frame}
\acro{NOMA}{non-orthogonal multiple access}
\acro{OMA}{orthogonal multiple access}
\acro{DFT}{discrete Fourier transform}
\acro{CDMA}{code-division multiple-access}
\acro{BCASC}{best complex antipodal spherical codes}
\acro{CBGC}{coherence-based Grassmannian codebook}
\acro{ICBP}{iterative collision-based packing}
\acro{MMSE}{minimum mean squared error}
\acro{MUSA}{multi-user shared access}
\acro{SIC}{successive interference cancellation}
\acro{SNR}{signal-to-noise ratio}
\acro{TDL-C}{tapped-delay-line-C}
\acro{LTE}{long-term evolution}
\acro{SINR}{signal-to-interference-plus-noise ratio}
\acro{SVD}{singular value decomposition}
\acro{KKT}{Karush-Kuhn-Tucker}
\acro{BLER}{block error ratio}
\acro{5G}{fifth-generation}
\acro{6G}{sixth-generation}
\acro{B5G}{beyond fifth-generation}
\acro{IoT}{internet-of-things}
\acro{PAPR}{peak-to-average-power ratio}
\acro{FFT}{fast-Fourier-transform}
\acro{IFFT}{inverse fast-Fourier-transform}
\acro{OFDM}{orthogonal frequency-division multiplexing}
\acro{BS}{base station}
\acro{UE}{user equipment}
\acro{MUD}{multiuser detection}
\acro{CWL}{codeword level}
\acro{MMSE}{minimum mean square error}
\acro{MRC}{maximum-ratio combining}
\acro{MF}{matched filter}
\acro{PIC}{parallel interference cancellation}
\acro{CRC}{cyclic-redundancy-check}
\acro{RB}{resource-block}
\acrodefplural{RB}{resource-blocks}
\acro{RMS}{root-mean-square}
\acro{DS}{delay spread}
\acro{LDPC}{low-density parity-check}
\acro{MIMO}{multiple-input multiple-output}
\acro{ITS}{intelligent transport systems}
\acro{V2X}{vehicle-to-everything}
\acro{V2V}{vehicle-to-vehicle}
\acro{V2I}{vehicle-to-infrastructure}
\acro{V2N}{vehicle-to-network}
\acro{V2P}{vehicle-to-pedestrian}
\acro{DSRC}{dedicated short-range communication}
\acro{C-V2X}{cellular-\ac{V2X}}
\acro{IEEE}{institute of electrical and electronics engineers}
\acro{MAC}{medium access control}
\acro{PHY}{physical}
\acro{CSMA}{carrier sense multiple access}
\acro{SB-SPS}{sensing-based semi-persistent scheduling}
\acro{5G-NR}{5th generation new-radio}
\acro{mMTC}{massive machine-type communication}
\acro{IGMA}{interleave-grid multiple access}
\acro{IDMA}{interleave-division multiple access}
\acro{ECDF}{empirical cumulative distribution function}
\acro{LLR}{log-likelhood-ratio}
\acro{IRS}{intelligent reflecting surface}
\acro{RIS}{reconfigurable intelligent surface}
\acro{LOS}{line-of-sight}
\acro{NLOS}{non-line-of-sight}
\acro{RV}{random variable}
\acro{CLT}{central limit theorem}
\acro{CDF}{cumulative distribution function}
\acro{IC}{interference cancellation}
\acro{SDP}{semidefinite programming}
\end{acronym}
\bibliographystyle{IEEEtran}
\bibliography{IEEEabrv,./LocalRefs}

\end{document}